\documentclass[11pt,manuscript]{aastex}
\input psfig.sty

\newcommand {\ea} {{\it et~al.}}
\newcommand {\be} {\begin{equation}}
\newcommand {\ee} {\end{equation}}

\shorttitle{GRBs from AC jets}

\shortauthors{Sikora \ea}

\begin{document}

\title{Gamma-ray bursts from alternating-current jets}

\author{Marek~Sikora\altaffilmark{1}$^,$\altaffilmark{2}, Mitchell C.~Begelman\altaffilmark{1}$^,$\altaffilmark{3}, 
Paolo~Coppi\altaffilmark{1}$^,$\altaffilmark{4}$^,$\altaffilmark{5} and Daniel~Proga\altaffilmark{1}$^,$\altaffilmark{6}}

\altaffiltext{1}{JILA, University of Colorado, Boulder, CO 80309-0440, USA}
\altaffiltext{2}{Permanent address: Nicolaus Copernicus Astronomical Center,
Bartycka 18, 00-716 Warsaw, Poland; \tt{sikora@camk.edu.pl}}
\altaffiltext{3}{also: Department of Astrophysical and Planetary Sciences, University of Colorado at Boulder; \tt{mitch@jila.colorado.edu}}
\altaffiltext{4}{JILA Visiting Fellow}
\altaffiltext{5}{Permanent address: Department of Astronomy, Yale University, P.O. Box 208101, New Haven, CT 06520-8101; \tt{coppi@astro.yale.edu}}
\altaffiltext{6}{\tt{proga@quixote.colorado.edu}}

\begin{abstract}

We propose a gamma-ray burst scenario involving relativistic jets dominated by Poynting flux with alternating toroidal magnetic fields.
Such a structure may arise naturally if the jet is formed and powered by
the accretion flow in the core of a collapsar. We conjecture that the
polarity of the toroidal magnetic field changes randomly due to
hydromagnetic turbulence driven by the magneto-rotational instability (MRI),
with the typical reversal time determined by the time scale for amplifying
magnetic fields up to a dynamically important level.  Poynting
flux-dominated jets with reversing B-fields provide a natural and efficient
way to dissipate energy via the reconnection process.
Gamma-rays are produced at the spatially separated reconnection sites. 
In this scenario, the emergent synchrotron radiation can be highly polarized and can form both smooth and spiky light curves.  
We note the possibility that cold and dense filaments can form in the
reconnection zones as the result of thermal instability. One could then
explain the production of very hard X-ray spectra, as due to bound-free absorption
of the synchrotron radiation. 

\end{abstract}


\section{INTRODUCTION}

Gamma-ray bursts (GRBs) probably result from the production of
ultra-relativistic jets in rapidly rotating, collapsing cores of massive
stars (Paczy\'nski 1993). The jets are presumably powered by energy extracted
electromagnetically from a rotating accretion disk and/or black hole.
 A remarkable fraction of the jet energy is
dissipated and re-radiated in the form of gamma-ray bursts, at distances
$\sim 10^{13}-10^{14}$ cm from the core.  The most popular models of energy
dissipation involve internal shocks, formed via collisions between
inhomogeneities propagating down the jet with different velocities (Rees \&
M\'esz\'aros 1994; Sari \& Piran 1995).  A very high level of velocity 
modulation at the source, of unknown origin, must be assumed in order to obtain an adequate rate of energy dissipation.  The dissipated energy is assumed
to be converted efficiently to relativistic electrons, which in turn
produce gamma rays by the synchrotron mechanism. These models predict
a spectral peak at photon energies consistent with observations 
(Zhang \& M\'esz\'aros 2002) and are relatively successful in explaining the
observed variety of light curves (Nakar \& Piran 2002).  However, they are
unable to explain the extreme hardness of X-ray spectra that are sometimes
observed (Ghirlanda et al. 2003).  

The difficulties encountered in modeling very hard X-ray spectra have 
motivated researchers to study other mechanisms of gamma-ray production. 
Proposed scenarios include thermal photospheric radiation, 
thermal multiple Compton scatterings, and Compton scattering by the bulk flow 
(Lazzati et al. 2000 and references therein). If the polarization measured by the RHESSI satelite 
(Coburn \& Boggs 2003) is real, even if overestimated, then the two former 
scenarios can be excluded. The bulk Compton mechanism can still work
(Begelman \& Sikora 1987; Eichler \& Levinson 2003; Lazzati et al. 2003), 
provided that the jet has an opening angle $\le 1/\Gamma$, where 
$\Gamma$ is the bulk Lorentz factor.

Another class of GRB models is based on the assumption that the jets are 
dominated by Poynting flux, even on very large scales 
(Lyutikov \& Blackman 2001; Spruit et al. 2001; Drenkhahn 2002).
In these models, energy is dissipated and particles are accelerated via 
reconnection of the magnetic field.  Particularly promising are models
in which the polarity of the toroidal magnetic field repeatedly flips.  Such a structure can result from flux injection by 
a rotating misaligned dipole (Spruit et al. 2001). In such ``striped-wind" models, which were originally proposed to model pulsars (Coroniti 1990),
reconnection occurs at the sites of field reversals.

Toroidal field reversals can also arise from the self-consistent evolution
of the magnetic field at the base of a jet, according to simulations that
follow MRI (magneto-rotational instability) driven accretion. The combination of shear, supported by 
the differential rotation of the accretion flow, with magnetic turbulence, 
driven by MRI (Proga et al. 2003), causes the polarity of the toroidal 
component to change with time (Proga et al., in preparation).  
Even if a weak poloidal field of given polarity is imposed at the outer 
boundary, the flow loses memory of this polarity by the time it reaches the
inner regions where the jet is formed. The field reversals are stochastic,
but the characteristic time scale corresponds to the time scale for winding
the toroidal field up to values at which the magnetic pressure becomes 
strong enough to power an outflow.  Consequently, a jet with magnetic 
reversals is formed.  A similar structure, albeit with a different origin,
was postulated for AGN jets by Lovelace et al. (1997). 

In this letter, we present our preliminary results on GRBs from jets with 
magnetic reversals. We state the model assumptions and flow 
parameters in \S{2.1}, and in \S{2.2} we discuss the production of
synchrotron radiation at the photosphere. In \S{2.3} we investigate the possible 
formation of sheets or filaments of cold, dense plasma and consider 
the hardening of the synchrotron spectrum by bound-free absorption.
In \S2.4 we make predictions about the basic features of the light curves, 
and in \S2.5 we show how X-ray flashes (XRFs) can be unified with GRBs in 
terms of our model. The main advantages of the model, as well as its 
uncertainties, are summarized in \S3.

\section{THE MODEL}

\subsection{Assumptions and basic parameters}
We assume that relativistic jets in GRBs are Poynting flux-dominated, and
are composed of magnetic domains with opposite polarities of the toroidal
magnetic field. In this model, GRBs are produced following the
annihilation/reconnection of magnetic fields at the boundaries between domains. The observed
gamma-ray emission comes from the region just downstream of the radius, $R_0$, where 
the jet becomes optically thin.

The jet is assumed to be conical and uniform in the transverse direction, and can be 
described by the following parameters:
the total energy, $E$; the ratio of the magnetic energy flux to 
the matter energy flux in the radiating region, $\sigma=L_B/L_M$; 
the bulk Lorentz factor, $\Gamma$; the opening angle of the jet, $\theta_j$; the GRB lifetime,
$t_{GRB}$; the characteristic width of a magnetic domain, $\lambda$; and the
ratio of electrons plus positrons to protons $f_e= n_e'/n_p'$.

In jets dominated by the toroidal magnetic component, 
\be L_B \simeq  2c u_B' \pi (R \theta_j \Gamma)^2 \,  , \label{LB} \ee 
where $u_B' \simeq {B_{\phi}'}^2/(8\pi)$ is the magnetic energy density in the jet comoving frame (quantities measured in the jet comoving frame
are primed, with the exception of the random electron Lorentz factor).  Combining this with the relation 
\be L_B = {E_B \over t_{GRB}} = {\sigma \over 1+\sigma} {E \over t_{GRB}}
\,  \label{LBE} \ee   
gives 
\be u_B' \simeq {1 \over 2 \pi c}\,{\sigma \over \sigma +1}\, {E\over t_{GRB}}\,
{1\over (\Gamma \theta_j)^2}\, {1\over R^2} \, . \label{uB1} \ee

Assuming $n_e' \ll (m_p/m_e) n_p'$, 
\be L_M \simeq  n_p' m_p c^3 (R \theta_j \Gamma)^2 \, , \label{LM} \ee
and noting that 
\be L_M \simeq {E_M \over t_{GRB}} = {1 \over \sigma +1} {E \over t_{GRB}}
\, , \label{LBM} \ee   
we obtain
\be n_e' = {1\over \pi m_p c^3} \, {n_e' \over n_p'} {1 \over  1+\sigma} \, 
{E\over t_{GRB}} \, {1 \over (\Gamma \theta_j)^2} \, {1\over R^2} \, .
\label{ne1} \ee

Hereafter, we adopt the following normalizations:
$t_{GRB} = 30 t_{30}$ s; $\Gamma = 100 \Gamma_2$;
$\theta_j = 0.1 \theta_{-1}$; $E= 10^{52} E_{52}$ erg; and 
$\lambda = 10^{10} \lambda_{10}$ cm.

\subsection{Synchrotron radiation at the photosphere}

If $c\,t_{GRB} \Gamma^2 > R$, the Thomson optical depth for radiation
produced 
within the flow at a distance $R$ and beamed within the Doppler cone is
\be \tau_T \simeq {R n_e' \sigma_T \over 2\Gamma} \, . \label{TT} \ee
Combining this with the eq. (\ref{ne1}) gives the distance of the photosphere
\be  R_0 \simeq 
7.9 \times 10^{13} \, 
f_e \sigma^{-1} \Gamma_2^{-3} \theta_{-1}^{-2} E_{52} t_{30}^{-1} \, {\rm cm} \, . \label{R0} \ee
Inserting eq. (\ref{R0}) into eq. (\ref{uB1}) gives 
\be u_{B,0}' = 
2.8 \times 10^9  \, f_e^{-2} \sigma^2 \Gamma_2^4 \theta_{-1}^2 E_{52}^{-1} t_{30}
  \, {\rm erg \ cm}^{-3}
\label{uB2}  \ee
and
\be B_0' = 2.7 \times 10^5  \,  f_e^{-1} 
\sigma \Gamma_2^2 \theta_{-1} E_{52}^{-1/2} t_{30}^{1/2} \, {\rm G} \, ,
\label{B}  \ee
while from inserting eq. (\ref{R0}) into eq. (\ref{ne1}) one gets
\be n_{e,0}' = 3.8 \times 10^{12}  \, f_e^{-1} 
\sigma \Gamma_2^4 \theta_{-1}^2 E_{52}^{-1} t_{30}   \, {\rm cm}^{-3} \, .
\label{ne}  \ee

Magnetic energy, released via the reconnection process, is transmitted to 
the plasma at a rate (per unit surface area)
\be F_{in}' = u_B' v_{in}' \ee
where $v_{in}'$ is the inflow velocity of magnetized plasma 
into the reconnection region.  Assuming that the dissipated energy is
shared equally among all particles dragged into the reconnection region, the electrons
reach Lorentz factor 
\be
\gamma_{inj}= {1 \over m_ec^2} {u_B' v_{in}' \over (n_e' + n_p') v_{in}'} 
= {\sigma (m_p/2m_e) \over 1 + f_e} 
\simeq 9.2 \times 10^2 f_e^{-1} \sigma 
\, . \label{gamma} \ee
These electrons produce synchrotron radiation peaked around
\be E_{p,0} = 
{2 h e \over 3 \pi  m_e c}  \gamma_{inj}^2 \xi_B B_0' \Gamma \simeq 
350 \, \xi_B f_e^{-3} \sigma^3 \Gamma_2^3 \theta_{-1} E_{52}^{-1/2} 
t_{30}^{1/2} \, {\rm keV} \, , \label{Epik} \ee
where $\xi_B = B_{rec}'/B' <1$ and $B_{rec}'$ is the intensity of 
the reconnected magnetic field. This formula put constraints on the value
of $\sigma$, which in our model is a free parameter. Noting that typical
spectral peak energies in GRBs are $\sim 300$ keV (Preece et al. 2000), one
can conclude 
that very high $\sigma$'s are allowed only for a high pair content 
($f_e \gg 1$). 
If the latter is determined by locally ({\it in situ}) produced pairs,
via absorption  of synchrotron self-Compton (SSC) photons by synchrotron 
photons, then $f_e$ is not
expected to be larger than a few. This is because $L_{SSC} / L_{syn}
\sim v_{in}'/(\xi_B c) \sim 1$ and pairs are produced in the Klein-Nishina
regime. This implies that GRB jets above the photosphere are only
mildly dominated by magnetic fields, i.e. $\sigma \sim$ a few.

Since the time scale of electron synchrotron energy losses,
\be t_{syn,0}' \simeq 
{3 m_e c^2 \over 4 c \sigma_T \xi_B^2 u_{B,0}' \gamma_{inj}} \simeq 2.4 \times 
10^{-5}  \, \xi_B^{-2} f_e^3 
\sigma^{-3} \Gamma_2^{-4} \theta_{-1}^{-2} E_{52}t_{30}^{-1} \, {\rm s} \, ,
\label{tsyn}  \ee
is about 6 orders of magnitude shorter than the dynamical time scale, 
$t_{dyn}' \sim (R/c)/\Gamma$, the electrons cool very rapidly. Steadily
injected into a uniform magnetic field, they produce a synchrotron spectrum
with a photon index $\alpha=-1.5$.  Such a spectrum is much softer than that observed 
in most GRBs (Preece et al. 2002). 

\subsection{Cold filaments and bound-free absorption?}  
Several mechanisms have been suggested to make the synchrotron spectra
harder, including synchrotron self-absorption (Papathanassiou 1999; Granot et al. 2000), very small pitch angle radiation (Lloyd \& Petrosian 2000; orginally elaborated by Epstein 1973), and jitter radiation (Medvedev 2001).  All of these mechanisms face a variety of difficulties (Ghisellini et al. 2000; Ghirlanda et al. 2003) and none is able to account the hardest spectra observed ($\alpha > 0$).  Synchrotron models of GRBs are therefore seriously challenged by
observations of extremely hard spectra.  There is, however, at least one way to rescue the synchrotron mechanism for GRBs: absorption of synchrotron
radiation by very dense and cold plasma sheets or  filaments.   Such filaments could be formed as follows:

Electrons cooling via synchrotron radiation and the SSC process, and then by thermal Comptonization of the synchrotron radiation, reach the Compton temperature $T_C' \sim 10^8 - 10^9$ K. Protons, very inefficient radiators and very weakly coupled to electrons via Coulomb interactions, might remain
mildly relativistic, losing energy only via adiabatic expansion.  However, the intense, small scale turbulence likely to be associated with 
reconnection could well drive plasma instabilities that couple the protons and electrons thermally on much shorter time scales (Begelman \& Chiueh 1988; 
Quataert \& Gruzinov 1999). In the latter case, energy would be drained from the protons to the electrons and radiated away (via thermal
Comptonization of synchrotron radiation) and protons would also reach the Compton  temperature. This cooling can be accompanied by
strong compression of the matter along the reconnected magnetic field lines.

Plasma at $T_C' \sim 10^8 - 10^9$ K is unstable to further cooling if the gas pressure is somewhat larger than the ambient radiation energy
density (Begelman \& McKee 1990).  This condition may be satisfied by a factor of a few, depending on the details of flow into the reconnection
zone.  If it is satisfied, plasma would cool down to temperatures $T_w' \sim 10^4$K and would form clouds or filaments with density
\be n_w' \sim u_B'/kT_w' \sim 3 \times 10^{21} f_e^{-2} \sigma^2 {\rm cm}^{-3} 
\, . \label{nw} \,  \ee 
Under such conditions a significant fraction of the gas is in a neutral state and bound-free absorption of synchrotron radiation 
can lead to significant hardening of the synchrotron spectrum.  Because of neutronization of the plasma 
in the central region and because only deuterium and $\alpha$-particles are
recovered by nucleosynthesis during the initial expansion (Derishev et
al. 1999; Beloborodov 2003), the absorber is expected to be strongly
dominated by HI, HeI and HeII.  It is only necessary for $\sim 1$\% of the
gas in the reconnection region to be in this state in order to provide
enough hardening of the synchrotron spectrum in the observed energy
ranges. Bound-free absorption hardens the 
spectrum of synchrotron radiation 
produced within a mist of a cold gas by $\Delta\alpha = 3$. If the
intrinsic synchrotron spectrum has $\alpha = -1.5$, this would lead to 
a photon index $\alpha = 1.5$.

\subsection{Light curves}

Gamma-ray pulses, produced by individual reconnection sheets, start to
build 
up at $R=R_0$ and, due to light travel time effects related
to the transverse size within the Doppler beam, reach a maximum after
\be t_{rise} \sim {R_0 \over 2c \Gamma^2} \simeq 0.13 \sigma^{-1}
t_{30}^{-1}
 \Gamma_2^{-5} \theta_{-1}^{-2} E_{52}   \, {\rm s} \, . \label{trise}
\ee

After the maximum the pulse decays, partly because the peak of the synchrotron spectrum moves to lower energies with increasing distance 
($\nu_{pk} \propto B' \propto 1/R$), and partly because of the decreasing efficiency of the reconnection process.  
(Note, that in the case of reconnection proceeding with a constant rate 
$\propto v_{in}' \simeq \xi_A v_A' \simeq \xi_A c$, the
bolometric flux would last $\sim \lambda/\xi_A c$ and would start to drop at a distance 
$\sim 10^{15} \lambda_{10} \xi_{A,-1}^{-1} \Gamma_2^2$ cm.)   Sequential pulses 
are separated by the magnetic reversal time, $t_{\lambda} \simeq \lambda/c$. 
The observed rise time and interval between pulses are equal for
\be \Gamma_c \simeq 85 \lambda_{10}^{-1/5} f_e^{1/5} \sigma^{-1/5} 
\theta_{-1}^{-2/5} E_{52} ^{1/5} t_{30}^{-1/5}
  \, . \label{Gammac} \ee
Thus, for $\Gamma > \Gamma_c$ the light curves are predicted to be
``spiky,"  
while for $\Gamma < \Gamma_c$ the light curves are expected to be smooth.  
This is assuming that all magnetic domains have similar widths. Since
we 
expect a broad distribution of domain sizes, the rise time, for a given 
Lorentz factor, should define a time 
variability filter, which suppresses variability on timescales shorter than
$t_{rise}$. The very strong dependence of the filter scale on the value of 
$\Gamma$ provides the new independent method for estimating the bulk
Lorentz factor.
 
\subsection{Total energetics and X-ray flashes}

As was recently pointed out (Frail et al. 2001; Bloom et al. 2003), the total energetics may be similar for all GRB, and the different luminosities can simply be due to different jet opening angles, i.e., $L \theta^2 \sim$ const.  This, combined with the well-established correlation 
$E_{p} \propto L^{1/2}$ (Lloyd et al. 2000; Yonetoku et al. 2003), suggests that $E_{p} \propto 1/\theta$.
Our model predicts $E_{p} \propto \Gamma^3 \theta$, and therefore can be reconciled with observations if 
\be \Gamma \propto \theta_j^{-2/3} \, , \label{gamatheta} \ee
assuming that in all GRBs $\sigma \sim $ a few. 

At present we are not able to verify this relation theoretically, however,
we can use it to make a prediction regarding the correlation between short-term variability and the value of
$E_{p}$. Namely, for larger $\theta_j$ and, therefore, lower luminosities and lower 
peak energies, we predict smaller bulk Lorentz factor, and this implies lower variability and smoother light curves (see previous section). The best 
objects to verify this prediction are XRFs, provided that they belong to the same class of phenomena as GRBs (Kippen et al. 2002).
In these objects $E_{p} < 30$ keV; therefore, XRFs having the same total energies as GRBs should be of much lower luminosity and 
produce much smoother light curves than typical GRBs.

\section{CONCLUSIONS}

Our main results can be summarized as follows:
\smallskip

\noindent
$\bullet$ Driven by MRI, accretion onto a black hole in a collapsar core may lead to the formation of jets with alternating toroidal magnetic fields;
\smallskip

\noindent
$\bullet$ Magnetic reconnection at the contact surfaces between domains with opposite magnetic polarities provide spatially separated particle acceleration regions. This provides a natural  source of short-term variability in GRBs, without the necessity of imposing rapid, high-amplitude modulations of the jet's density and speed at its base;
\smallskip

\noindent
$\bullet$ Under some circumstances, a portion of the plasma in the reconnection sites can condense into a very cold, dense phase. A mist of cold 
clouds/filaments can harden the synchrotron spectrum via bound-free absorption;
\smallskip

\noindent
$\bullet$ Our scenario predicts the formation of both smooth and spiky light curves, and provides a method to estimate the bulk Lorentz factor;
\smallskip

\noindent
$\bullet$ The model can be reconciled with observed correlations between variability, luminosity and spectral peak energy, provided that the bulk
Lorentz factor anti-correlates with the jet opening angle.  This allows us to make a prediction that the less luminous objects
should have smoother light curves. In particular, this prediction applies to XRFs, provided that they represent the same phenomenon as GRBs. 
\smallskip

The main uncertainties of the proposed scenario concern the efficiency of 
the magnetic reconnection process and the formation of cold and dense gas
condensations. Unfortunately, the reconnection process is still not well
understood, especially under the extreme conditions of relativistic
outflows (Kirk \& Skj{\ae}raasen 2003), and with rapidly cooling plasmas.  
The possible occurrence of thermal instability depends critically
on such unknowns as the strength and geometry of the magnetic field in 
the cooling region and the efficiency of energy transport from protons to 
electrons.

Furthermore, we did not follow the role of pairs in detail. Our 
preliminary estimates show that SSC photons are effectively absorbed by 
synchrotron photons, to produce pairs within the reconnection sheets as
well as outside. How they will affect the shape of the synchrotron spectrum, and
its evolution, is not clear yet.  Modeling of pair cascades is also
important to provide constraints on the magnetization parameter $\sigma$. 

Finally, we note that the creation of cold plasma condensations
in our scenario can also provide a basis for bulk Compton models of GRBs. We 
recall that the plasma is required to be cold in this model in order to 
provide high linear polarization  (Poutanen 1994). It must also satisfy another condition,
$\theta_j < 1/\Gamma$ (Begelman \& Sikora 1987; Lazzati et al. 2003), the same as for internal shock models (Nakar et al. 2003).   
Confirmation of high polarization in GRBs, in particular in less luminous  
objects in which presumably $\theta_j \gg 1/\Gamma$, would be a critical step toward discriminating between the reconnection model and others 
(Lyutikov et al. 2003).

\acknowledgments

The project was partially supported by Polish KBN grants 5 P03D 00221,
PBZ-KBN-054/P03/2001 and NSF grant AST-0307502.
D.P. acknowledges support from NASA LTSA grant NAG5-11736.
 M.S. and P.C. thank the Fellows of JILA, University 
of Colorado, for their hospitality.

\end{document}